\documentclass[fleqn,12pt,twoside]{article}
\usepackage{espcrc1}

\usepackage{graphicx}
\usepackage[figuresright]{rotating}
\def\lp {\left( }
\def\rp {\right) }
\def\lb {\left[ }
\def\rb {\right] }

\def\ni {\noindent}
\def\nn {\nonumber}

\def\rar {\rightarrow}
\def\lrar {\leftrightarrow}

\def\beq{\begin{equation}}
\def\eeq{\end{equation}}
\def\bea{\begin{eqnarray}}
\def\eea{\end{eqnarray}}

\def\cI {{\cal{I}}}

\def\cO{{\cal{O}}}
\def\cT {{\cal{T}}}

\def\a{\alpha}
\def\b{\beta}

\def\d{\delta}

\def\e{\epsilon}

\def\m{\mu}

\def\O {\Omega}
\def\p {\pi}
\def\r{\rho}
\def\s{\sigma}

\def\ub {\bar u}

\def\sp {\!+\!}
\def\sm {\!-\!}
\def\st {\!\times \!}
\def\cd {\!\cdot\!}

\def\bk {\mbox{\boldmath $k$}}

\def\br {\mbox{\boldmath $r$}}

\def\bro {\mbox{\boldmath $\rho$}}

\def\bsig {\mbox{\boldmath $\sigma$}}
\def\btau {\mbox{\boldmath $\tau$}}

\def\bnb {\mbox{\boldmath $\nabla$}}
\def\bO {\mbox{\boldmath $\Omega$}}
\def\bp {\mbox{\boldmath $p$}}

\def\bq {\mbox{\boldmath $q$}}
\def\bQ {\mbox{\boldmath $Q$}}

\newcommand{\AmS}{{\protect\the\textfont2
  A\kern-.1667em\lower.5ex\hbox{M}\kern-.125emS}}

\title{Two and Three Nucleon Forces}

\author{M. R. Robilotta\\[2mm]
{Instituto de F\'{\i}sica, Universidade de S\~{a}o Paulo,\\
C.P. 66318, 05315-970, S\~{a}o Paulo, SP, Brazil\\
E-mail: robilotta@if.usp.br
}}
       
\begin{document}

\maketitle

\begin{abstract}
Chiral symmetry allows two and three nucleon forces to be treated in a single 
theoretical framework. 
We discuss two new features of this research programme at $\cO(q^4)$  
and the consistency of the overall chiral picture.
\end{abstract}

\section{CHIRAL SYMMETRY}
The venerable idea that nuclear forces are due to pion exchanges indicates 
that processes involving different number of nucleons are related, owing to
the common presence of some basic subamplitudes describing either single 
($N \rar \p N$) or multipion 
($\p \p \rar \p \p$, $\p N \rar \p N$, $\p N \rar \p \p N$, ...) interactions.
As the latter class encompasses free cases, relationships with scattering 
data are also possible.
Nowadays, this web of interconnections can be explored consistently by means of 
effective chiral lagrangians, in which just pions and nucleons are treated 
as explicit degrees of freedom.
The rationale for this approach is that the quarks $u$ and $d$, which have small 
masses,
dominate low-energy interactions.
One then works with a two-flavor version of QCD and treats these masses 
as perturbations in a chiral symmetric lagrangian.
Quark mass contributions are included systematically by means of a chiral 
perturbation 
theory $(ChPT)$, which allows the relevant dynamical features of QCD 
to be properly incorporated into the nuclear force problem.
Kinematical constraints imposed over a given subamplitude by the number of nucleons 
present in the system are automatically taken into account by the 
use of field theory techniques.

In order to perform chiral expansions, one uses a typical scale $q$, 
set by either pion four-momenta or nucleon three-momenta, such that $q<<1$ GeV.
As far as the nuclear force problem is concerned, one notes that the free $\p N$ 
amplitude begins at $\cO(q)$ and two chiral expansions 
up to $\cO(q^4)$ are presently available.
One of them employs the so called heavy baryon approximation\cite{FetM},
whereas the other one is fully covariant\cite{BL}. 
In the case of the $NN$ potential, the $OPEP$ provides the leading contribution, 
which 
begins\cite{Bira} at $\cO(q^0)$.
The more complex two-pion exchange potential $(TPEP)$ begins at $O(q^2)$ and
there are two independent expansions up to $O(q^4)$ in the literature, 
based on either heavy baryon\cite{HB} or covariant\cite{HR,HRR} ChPT.
The leading contribution to the three-nucleon force is associated with two-pion
exchange and begins at $\cO(q^3)$.
Quite generally, asymptotic (large $r$) expressions for the various potentials 
have the status of theorems and can be written in the form
 $\cO(q^L)\lb 1 + \cO(q) + \cO(q^2) + \cdots\rb$, where $L$ is the leading order.
This structure in terms chiral layers has little model dependence.

\section{TWO-NUCLEON POTENTIAL: TWO-PION EXCHANGE}

In the last fifteen years, the systematic use of chiral symmetry led to a 
considerable improvement in the understanding of $TPEP$ dynamics.
Here we briefly describe the problem, in a perspective biased by the work
done by our group\cite{HR,HRR}.
At $\cO(q^4)$, the dynamical content of the $TPEP$ is given by three 
families of diagrams, shown in the figure below.
Family I corresponds to the minimal realization of chiral symmetry and begins 
at $\cO(q^2)$, whereas family II is $\cO(q^4)$ and associated with 
pion-pion correlations. 
Both of them depend only on the constants $g_A$ and $ f_\pi$.
Family III begins at $\cO(q^3)$ and depends on low energy constants
(LECs, represented by the black dots), which can be extracted from either $N N$ or 
$\pi N$ scattering data.
In the latter case, the calculated $TPEP$ becomes a theoretical prediction.

\vspace{3mm}

\begin{center}
\begin{tabular}{@{}c}
\includegraphics[width=20pc]{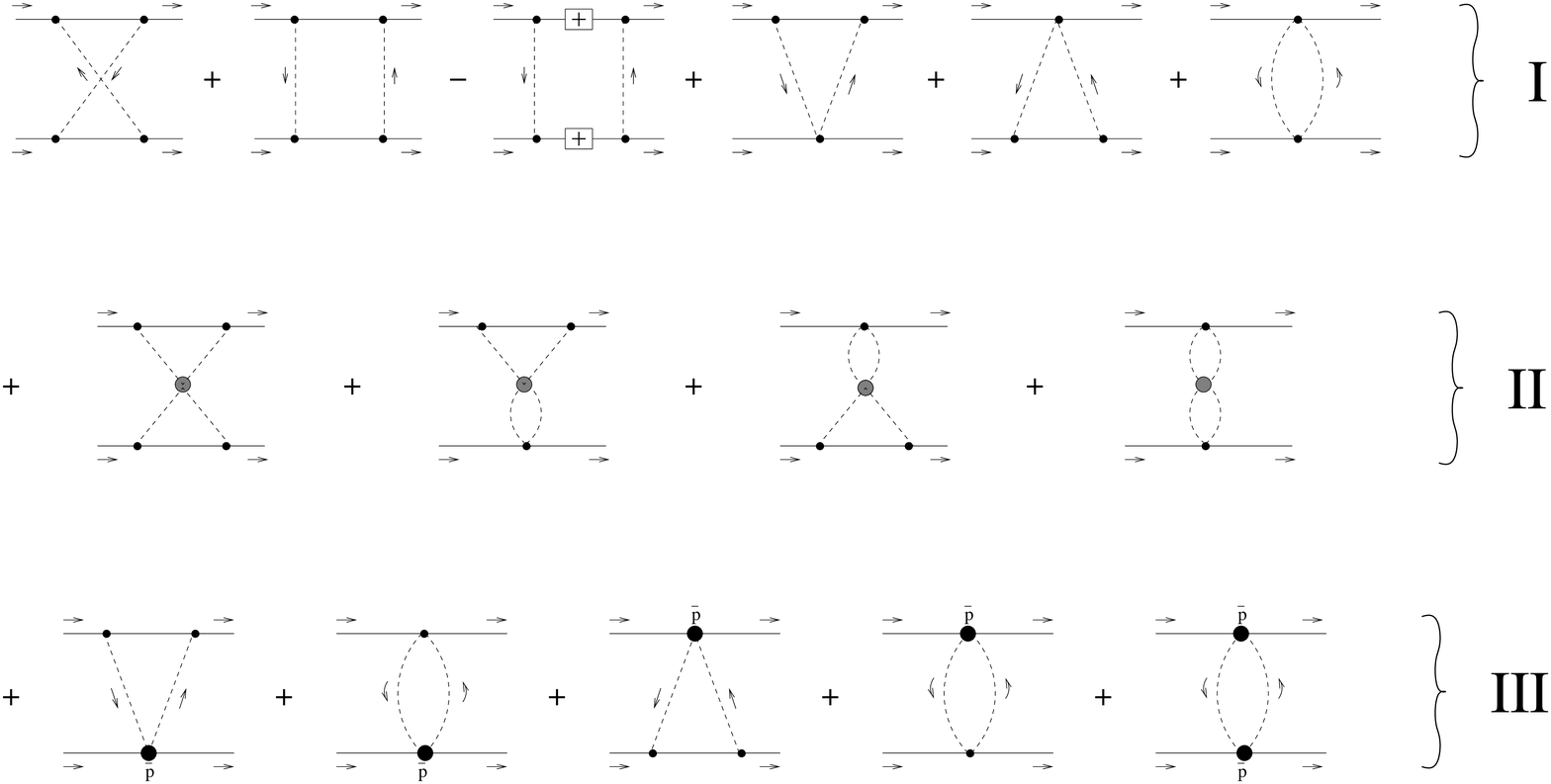}
\end{tabular}
\end{center}

Relativistic effects arising from the covariant treatment of loop integrals
are present even when the external nucleon momenta are small
and determine the form of asymptotic chiral theorems.
On the other hand, these effects are numerically small at distances of 
physical interest.
The chiral picture is well supported by empirical scattering data.

\begin{center}
\begin{tabular}{@{}l}
\includegraphics[width=18pc,angle=0]{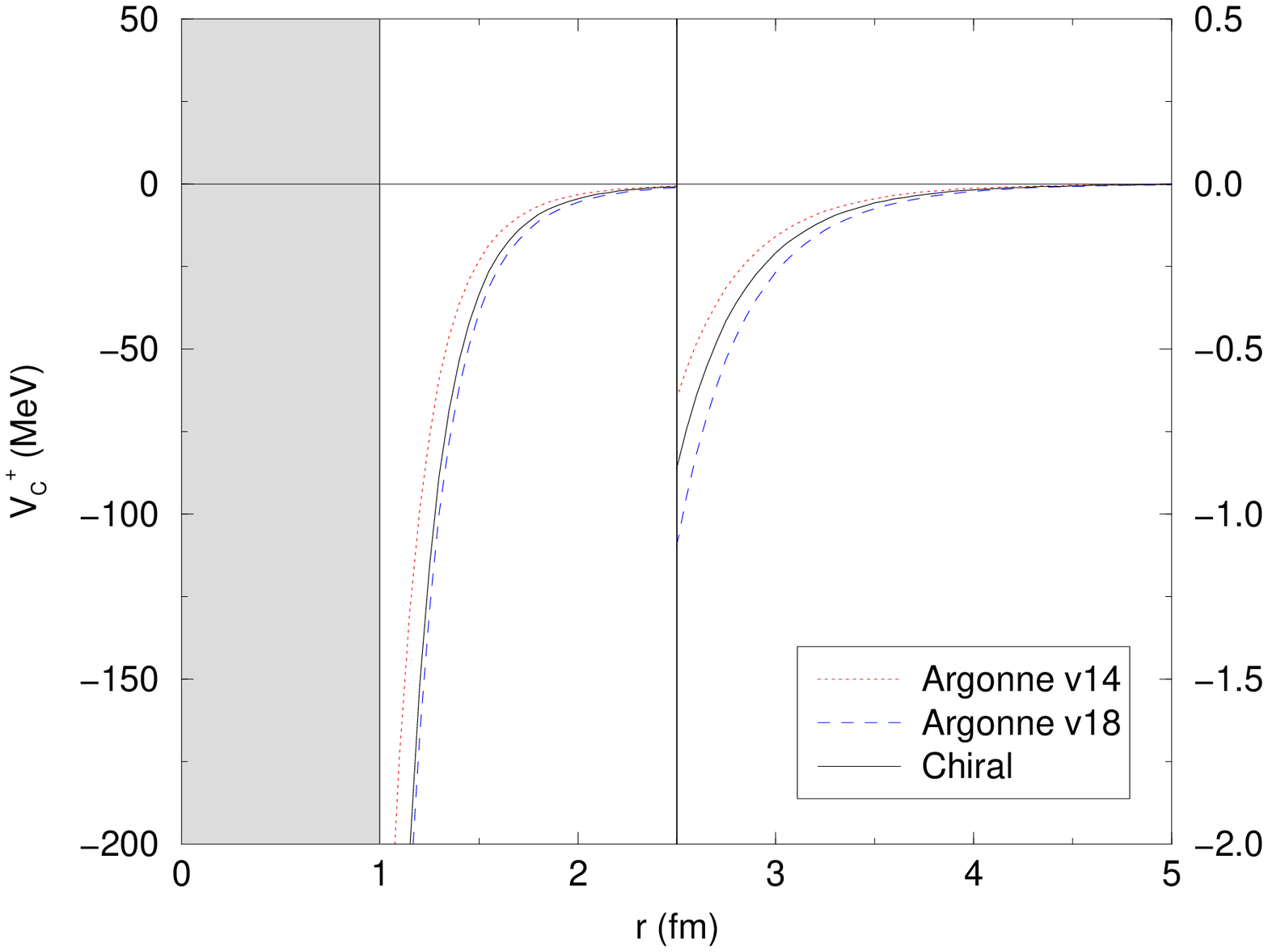}
\hspace{6mm}
\includegraphics[width=18pc,angle=0]{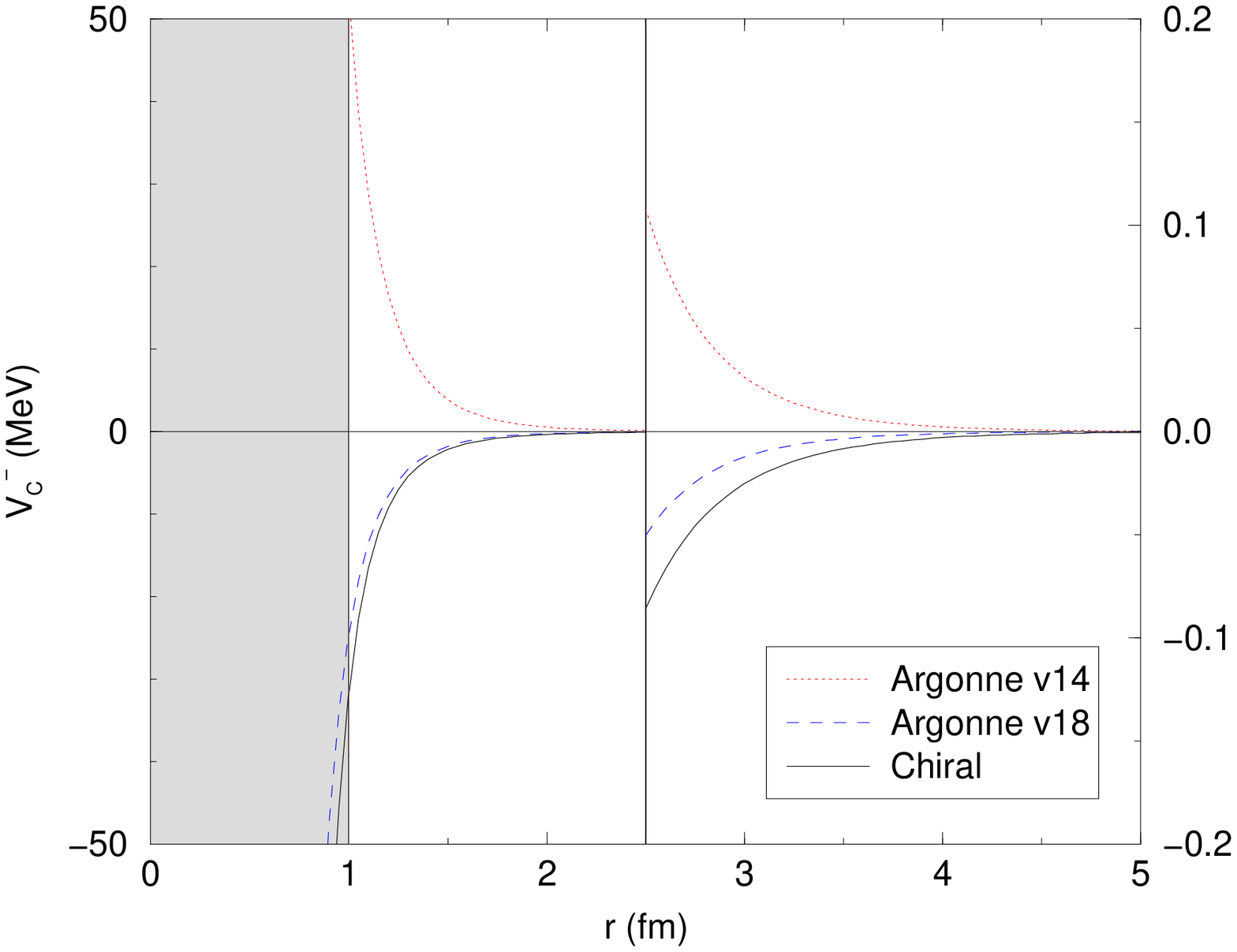}
\end{tabular}
\end{center}

As far as dynamics is concerned\cite{HRR}, family I strongly dominates the components
$V_{LS}^+$, $V_T^+$, $V_{SS}^+$ and $V_C^-$, whereas family III accounts almost 
entirely for $V_C^+$, $V_T^-$and $V_{SS}^-$.
Contributions from family II are rather small.
In the figures above we show the isospin even and odd central components,
which begin respectively at $\cO(q^3)$ and $\cO(q^2)$.
It is interesting to note that this hierarchy is not supported by the figures.

\section{DRIFT EFFECTS}

In the rest frame of a many-body system, used in the calculation of its static and 
scattering properties, the center of mass of a two-body subsystem is allowed to drift.
In the case of a three-body system, internal interactions are described by the function 

\vspace{-5mm}

\bea
&& W(\br', \bro'; \br, \bro)  = -\, \frac{1}{(2\p)^{12}}\; \lb 2/\sqrt{3} \,\rb^6
\int d\bQ_r \, d\bQ_\r \, d\bq_r \, d\bq_\r \;
\nn\\[1mm]
&& \;\;\;\;\; \;\;\;\;\; \times\;
\e^{i \lb \bQ_r \cdot (\br' \sm \br) + \; \bQ_\r \cdot (\bro' \sm \bro)  
+ \bq_r \cdot (\br' \sp \br) /2 + \bq_\r \cdot (\bro' \sp \bro) /2   \rb}
\;\; \bar{t}_3 (\bQ_r, \bQ_\r, \bq_r, \bq_\r)  \;,
\nn
\eea

\ni
where $\bar{t}_3$ is the proper part of the non-relativistic three-body transition
matrix, $\br$ and $\bro$ are usual Jacobi variables, and $\bQ_i=(\bp'_i \sp \bp_i)/2$,
$\bq_i=(\bp'_i \sm \bp_i)$, for $i=(r,\r)$.
In this framework, two-body interactions between nucleons $1$ and $2$ correspond to 
$\bq_\r = 0 $ and are described by 
  
\vspace{-6mm}

\bea
\bar{t}_3 (\bQ_r, \bQ_\r, \bq_r, \bq_\r) 
= (2\p)^3 \, \lb \sqrt{3}/2 \rb^3 \, \d^3(\bq_\r) \; 
\bar{t}_2(\bQ_\r, \bQ_r, \bq_r) \;.
\nn
\eea

The important feature of this result is that $\bar{t}_2$, the two-body $t$-matrix, 
depends on the variable $\bQ_\r$, which incorporates drift effects into the problem.
In the center of mass of the two-body subsystem one has $\bQ_\r=0$ and finds,
for each isospin channel $(\pm)$, the usual spin structure, given by\cite{HR} 

\vspace{-5mm}

\bea
\left. \bar{t}_2^\pm \rb _{cm}  
=  t_C^\pm  + \frac{\bO_{LS}}{m^2} \, t_{LS}^\pm 
+ \frac{\bO_{SS}}{m^2}\, t_{SS}^\pm + \frac{\bO_{T}}{m^2}\, t_{T}^\pm 
+ \frac{\bO_{Q}}{m^4}\, t_{Q}^\pm \;,
\nn
\eea

\ni
where the $\bO_i$ are spin operators.
In this case, the two-body interaction does not depend on $\bQ_\r$ and is 
completely decoupled from the larger system it is immersed in.
Corrections due to the motion of the two-body center of mass can be derived by 
evaluating the covariant scattering amplitude $\cT_2$ in the rest frame of the
three-body system and expressing the result in terms of two-component spinors. 
As this amplitude contains no approximations, all terms involving 
the variable $\bQ_\r$ can be interpreted as drift effects.
In the spirit of chiral perturbation theory, one next expands the amplitude in 
a power series and truncates it at a given order. 
In configuration space, the variables $\bQ_r$ and $\bQ_\r$ correspond to 
non-local operators, associated with gradients acting on the wave function.
In order to restrict the corresponding complications to a minimum, 
we consider only linear terms in these momenta
({\em linear gradient approximation}).

Explicit inspection of the $OPEP$ indicates that it has a rich drift structure which,
however, lies beyond the linear gradient approximation\cite{Rdrift}.
In the case of the $TPEP$, our $\cO(q^4)$ covariant amplitudes\cite{HR} have the 
general form  

\vspace{-5mm}

\bea
&& \cT^\pm = \sum_{\a,\b} \; \cI_{\a\b} \;\;
[\ub\,\Gamma_\a \, u]^{(1)} \,  
[\ub\, \Gamma_\b \,u]^{(2)} \;,
\nn
\eea

\vspace{-2mm}

\ni
where the $\cI_{\a\b}$ and $\Gamma_i$ are respectively Lorentz scalar amplitudes 
and Dirac spin operators.  
In the linear gradient approximation, the functions $\cI_{\a\b}$ do not depart from 
their
center of mass values and the only sources of drift corrections are the spin 
functions. 
These were studied in ref.\cite{Rdrift} and give rise to the structure 

\vspace{-7mm}

\bea
&& t_2^\pm  =  \left. t_2^\pm\rb_{cm}  + \frac{\bO_D}{m^2}\, t_D^\pm
\;\;\;\;\;\;\;\; \lrar \;\;\;\;\;\;\;\;
\bO_D = i\, (\bsig^{(1)} \sm \bsig^{(2)}) \cd \bq_r \st \bQ_\r /2\sqrt{3}\;.
\nn
\eea

\vspace{-2mm}

This $\cO(q^4)$ result springs directly from Lorentz covariance and is model 
independent.
Its Fourier transform yields the configuration space structure 

\vspace{-7mm}

\bea
&& V(r)^\pm  =  \left. V(r)^\pm \rb_{cm} + V_D^\pm \, \O_{D}
\;\;\;\;\;\;\;\; \lrar \;\;\;\;\;\;\;\;
\O_D = \frac{1}{4\sqrt{3}}\, (\bsig^{(1)}\sm \bsig^{(2)}) \cd \br \st \;,
(-i \bnb^{^{\!\!\!\!\!\!\!\!^\leftrightarrow}}_\r)
\nn\\
&& V_D^{\pm}(r)  =  \frac{\mu^2}{m^2}\,\frac{1}{x}\, \frac{d}{d x}\,U_D^{\pm}(x)
\;\;\;\;\;\;\;\; \lrar \;\;\;\;\;\;\;\;
U_D^\pm(x) =  - \int \frac{d \bq_r}{(2\pi)^3}\, e^{i\,\bq_r \cdot \br }\; 
t_D^\pm(q_r) \;,
\nn
\eea

\vspace{-2mm}

\ni
where $x=m_\p r$ and
$\bnb^{^{\!\!\!\!\!\!\!\!^\leftrightarrow}}_\r
= \bnb^{^{\!\!\!\!\!\!\!\!^\rightarrow}}_\r
- \bnb^{^{\!\!\!\!\!\!\!\!^\leftarrow}}_\r $.
It is worth noting that this atisymmetric form for the spin operator, already 
used in refs.\cite{BDK}, indicates that the drift potential enhances the 
role of $P$ waves in trinuclei.

\vspace{18mm}
\begin{center}
\begin{tabular}{@{}l}
\hspace*{-6mm}
\includegraphics[width=20pc,angle=0]{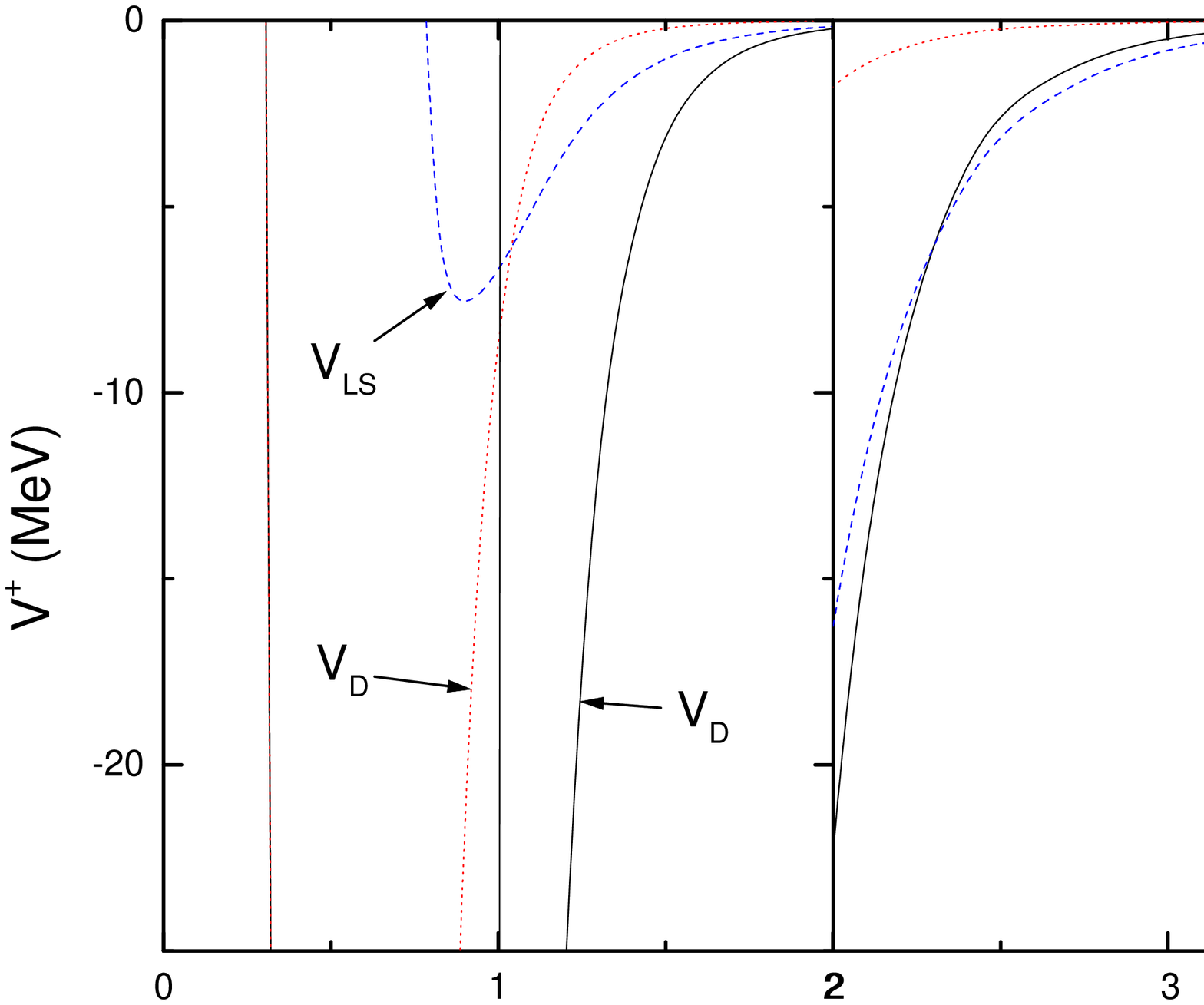}
\hspace*{-4mm}
\includegraphics[width=20pc,angle=0]{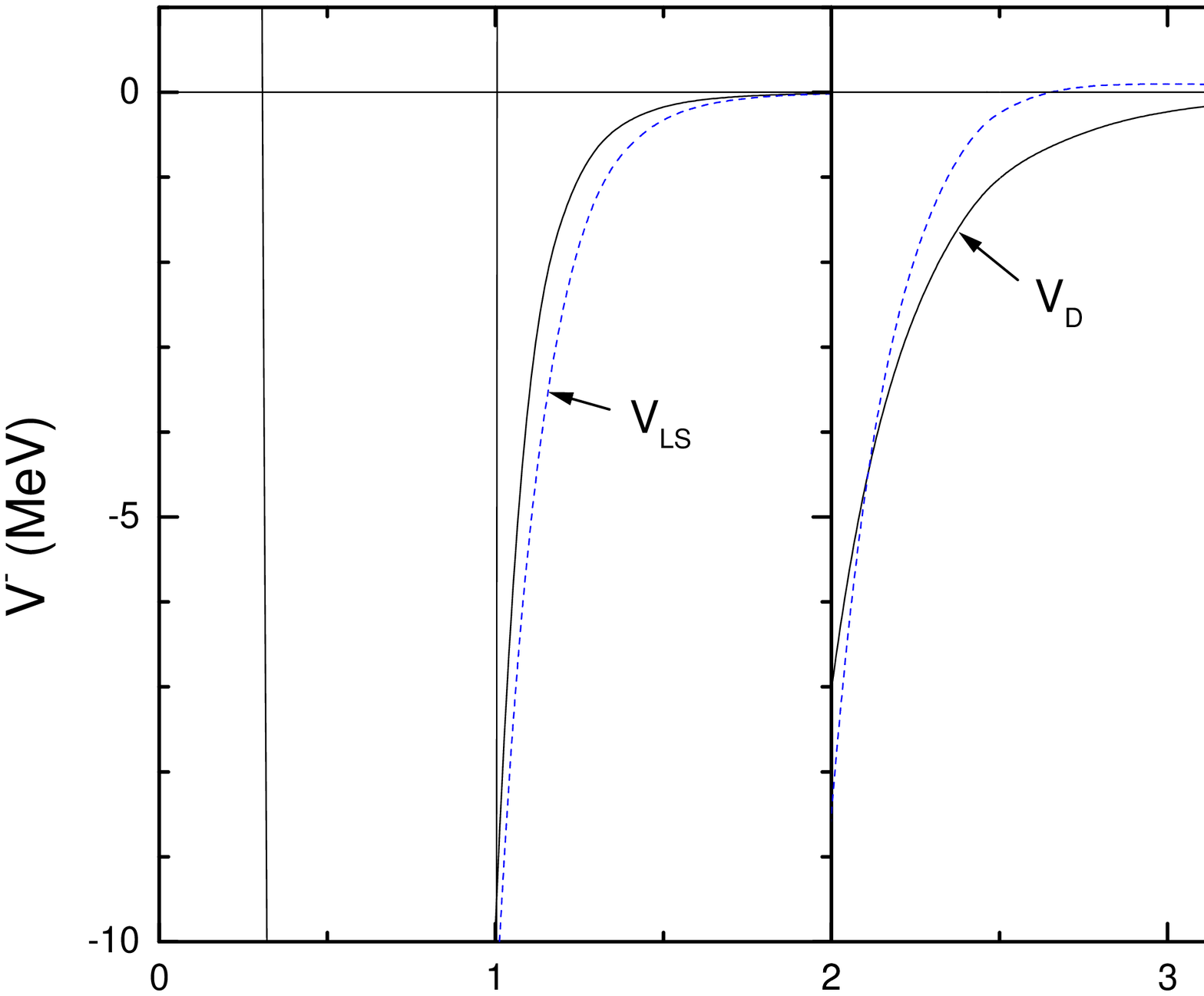}
\end{tabular}
\end{center}

\vspace{-30mm}

The figures above display the profile functions for the drift and spin-orbit 
potentials derived from our $\cO(q^4)$ expansion of the $TPEP$.
They do not include short range effects and cannot be trusted for $r< 1$fm.
As far as chiral symmetry is concerned, drift corrections begin at $\cO(q^4)$ and, 
in principle, should be smaller than the spin-orbit terms, which begin at $\cO(q^3)$.
However, in the isospin even channel, this chiral hierarchy is not respected and
one may expect important drift effects.
Finally, it is important to stress that the origin of drift effects is kinematical,
and not dynamical.

\section{TWO-PION EXCHANGE THREE-NUCLEON POTENTIAL}

The leading term in the three-nucleon potential has the longest possible range and 
corresponds to the process known as $TPE \sm 3NP$, in which a pion emitted by one of 
the nucleons is scattered by a second one, before being absorbed by the third nucleon.
It is closely related to the $\p N$ scattering amplitude, 
which is $\cO(q)$ for free pions and becomes $\cO(q^2)$ within the 
three-nucleon system.
As a consequence, the three-body force begins at $\cO(q^3)$.
This leading component has been available since long\cite{TM,CDR}.
The extension of the chiral series to $\cO(q^4)$ requires the inclusion of single
loop effects and is associated with a large number of diagrams.
Here, we concentrate on the particular set of processes which belong to the 
$TPE \sm 3NP$ class. 

Quite generally, the connected part of the  non-relativistic two-pion exchange 
amplitude can be written as 

\vspace{-6mm}

\bea
\bar{t}_3 = \frac{g_A^2}{4f_\p^2}\,
\frac{\bsig^{(1)}\cd \bk}{\bk^2 \sp \m^2}\;
\frac{\bsig^{(2)}\cd \bk'}{\bk^{'2}\sp \m^2}\;
\lb \btau^{(1)}\cd \btau^{(2)}\, D^+ 
- i\, \btau^{(1)}\times \btau^{(2)} \cd \btau^{(3)} 
\lp D^- + \frac{i}{2  m}\, \bsig^{(3)} \cd \bk' \st \bk \; B^- \rp \rb \;,
\nn
\eea

\ni 
where $g_A$ and $f_\p$ represent the axial nucleon and pion decay constants.
The subamplitudes $D^\pm$ and $B^\pm$ carry the dynamical content of 
the $\p N$ interaction and receive contributions from both tree diagrams and loops.
Their chiral content has been discussed, in covariant ChPT, by Becher and 
Leutwyler\cite{BL}.
The amplitude $\bar{t}_3$ corresponds to a configuration space potential of the form 

\vspace{-6mm}

\bea
V_3(\br, \bro) =  \btau^{(1)}\cd \btau^{(2)} \; V_3^+(\br, \bro) 
+ \btau^{(1)}\times \btau^{(2)} \cd \btau^{(3)} \; V_3^-(\br, \bro) 
+ \mathrm{cyclic \; permutations}\;.
\nn
\eea

\vspace{-1mm}

The inclusion of $\cO(q^4)$ contributions gives rise to both numerical corrections
in pre-existing strength coefficients $(C_i^\pm)$ and new structures in the profile 
functions.
Schematically, one has\cite{Rpipiex} 

\vspace{-6mm}

\bea
V_3^+(\br, \bro) &\!\!=\!\!& C_1^+ \lb \mathrm{\, old \; term\,} \rb
+ C_2^+ \lb \mathrm{\, old \; term \,} \rb
+ C_3^+ \lb \mathrm{\, gradients \; acting \; on \; loop \; function \,} \rb \;,
\nn\\
V_3^-(\br, \bro)  &\!\!=\!\!& C_1^- \lb \mathrm{\, old \; term \,} \rb
+ C_2^- \lb \mathrm{\, gradients \; acting \; on \; loop \; function \,} \rb 
\nn\\
&\!\! +\!\!& C_3^- \lb \mathrm{\, non\sm local \; terms \,} \rb \;.
\nn
\eea

\vspace{-1mm}

In this result, the loop functions are given by Fourier transforms of pion 
propagators multiplying Feynman integrals and the non-local terms 
are linear in gradients acting on the wave function.
The strength coefficients of the potential depend on well determined parameters
$(m_N, m_\p, g_A, f_\p)$, on the value of the scalar form 
factor at the Cheng-Dashen point ($\s_{CD}$) and on the LECs $c_3$ and $c_4$.
Adopting $\s_{CD}=60$ MeV and extracting the LECs from the $\p N$ subthreshold 
coefficients $d_{01}^+ = 1.14 \, m_\p^{-3}$ and  
$b_{00}^- = 10.36 \, m_\p^{-2}$, one finds the results quoted 
in the table below.

\begin{center}
\begin{tabular} {|c|c|c|c|c|c|c|}
\hline
	& $C_1^+$ & $C_2^+$ & $C_3^+$ & $C_1^-$ & $C_2^-$ & $C_3^-$  \\ \hline
$\cO(q^3)+\cO(q^4)$	 & 0.794 	& -2.118	& 0.011	& 0.691	&-0.025 & 0.021	\\ \hline
Brazil \cite{CDR}	& 0.920 & -1.99	& -	& 	0.67	& -	& -	\\ \hline
\end{tabular}
\label{T2}
\end{center}

The relatively small changes in these parameters are due to the use of ChPT.
At the chiral order one is working here, new effects associated 
with both non-local interactions and loop effects begin to show up.
They correspond to the terms proportional to the parameters 
$C_3^+$, $C_2^-$ and $C_3^-$.

\section{THE CHIRAL PICTURE}

The systematic application of chiral symmetry to the study of nuclear forces gives 
rise to a picture in which the various effects begin to appear at different orders.
This is indicated in the table below, which also includes the drift potential.

\begin{center}
\begin{tabular} {|c|lll|}
\hline
beginning	  			& TWO-BODY 		 & TWO-BODY		 & THREE-BODY 	\\ \hline
$\cO(q^0)$		& OPEP: $V_T^-, V_{SS}^-$	&& 			   	  		  	 	 \\ \hline
$\cO(q^2)$		& OPEP: $V_D^- \;\;\;\;\;$ & TPEP: $V_C^-; V_T^+, V_{SS}^+$ &\\ \hline
$\cO(q^3)$		&& TPEP: $V_{LS}^-, V_T^-, V_{SS}^-; V_C^+, V_{LS}^+$ 
& TPEP: $C_1^-; C_1^+, C_2^+$ 	   		  				   			 		 \\ \hline
$\cO(q^4)$		&& TPEP: $V_D^-; V_Q^+, V_D^+$ & TPEP: $C_2^-; C_3^-, C_3^+$\\ \hline
\end{tabular}
\label{T2}
\end{center}

The chiral series has been tested, by assessing the relative importance of 
$O(q^2)$, $O(q^3)$ and $O(q^4)$ terms in each component of the $TPEP$\cite{HR}.
One finds satisfactory convergence at distances of physical interest, 
except for $V_C^+$, where the ratio between $\cO(q^4)$ and $\cO(q^3)$ contributions 
is larger than $0.5$ for distances smaller than $2.5$ fm.

Chiral perturbation theory also predicts relative sizes for the various 
dynamical effects. 
For instance, it allows one to expect that $V_C^-$ should be larger than $V_C^+$,
since these terms begin respectively at $\cO(q^2)$ and $\cO(q^3)$.
Empirical data defy this expectation and, in the figures of section 2,
it is possible to note that $V_C^+$ is about 10 times
larger than $V_C^-$ at $2.5$ fm.
The same pattern is followed by the drift potential, which begins at $\cO(q^4)$ and, 
in principle, should be smaller than the spin-orbit terms, which begin at $\cO(q^3)$.
However, in the isospin even channel, the same dynamical contribution 
that operates in the central potential subverts the expected chiral hierarchy,
as shown in the figures of section 3.

The enhancement of some interactions in the isoscalar sector is a 
puzzling aspect of the chiral picture.
The numerical reasons for this behavior can be traced back to the large sizes 
of some of the LECs used in the calculation, which are dynamically generated 
by processes involving delta intermediate states.
Therefore the explicit inclusion of delta degrees of freedom in a covariant 
calculation could shed light into this problem.
On the other hand, one should also bear in mind that the very use of perturbation 
theory may not be suited to describe isoscalar interactions in regions of 
physical interest.
This is suggested by the study of the nucleon scalar form factor\cite{delta}, 
which is closely related with the central $NN$ interaction.
It indicates that the chiral angle is not small around the nucleon and the 
expected chiral hierarchy between nucleon and
delta contributions is already subverted around 1.2 fm.


\end{document}